\documentclass[aps,nofootinbib,prd,eqsecnum,showpacs,showkeys,preprintnumbers]{revtex4-2}
\usepackage[caption=false]{subfig}
\usepackage{graphicx}
\usepackage{amsmath}
\usepackage{amsfonts}
\usepackage{amssymb}
\usepackage{float}
\usepackage{color}
\usepackage{bm}
\usepackage{mathrsfs}
\usepackage{epstopdf}
\usepackage{url}
\usepackage{footnote}
\usepackage{textcomp}
\usepackage{ulem}
\usepackage{esint}
\usepackage[unicode=true, pdfusetitle,
 bookmarks=true,bookmarksnumbered=false,bookmarksopen=false,
 breaklinks=false,pdfborder={0 0 1},backref=false,colorlinks=false]{hyperref}
\usepackage{multirow}
\usepackage{pifont}

\begin{document}

\title{Dark matter via Baryogenesis: Affleck-Dine Mechanism in the Minimal
Supersymmetric Standard Model}
\author{K. El Bourakadi$^{1,2}$}
\email{k.elbourakadi@yahoo.com}
\author{M. Ferricha-Alami$^{1,2}$}
\email{ferrichaalami@yahoo.fr}
\author{Z. Sakhi$^{1,2}$}
\email{zb.sakhi@gmail.com}
\author{M. Bennai$^{1,2}$}
\email{mdbennai@yahoo.fr}
\author{H. Chakir$^{1,2}$}
\email{chakir10@gmail.com }
\date{\today }

\begin{abstract}
We conducted an investigation into Affleck-Dine baryogenesis within the
context of D-term inflation, specifically focusing on its relationship with a
recent reheating formalism. It was found that by considering a specific
reheating temperature, the observed baryon asymmetry can be accounted through Affleck-Dine baryogenesis. Additionally, the majority of gravitinos
are inferred to be generated from the decay of the next-to-lightest
supersymmetric particle, with Q-balls potentially serving as a source of
gravitinos via NSP decay. The temperature at which decay occurs depends on
the charge of the Q-balls, which is determined by the fragmentation of the
Affleck-Dine condensate. Remarkably, the gravitino mass required for dark
matter aligns naturally with the theoretical gravitino mass.
\end{abstract}

%\author{M. Ferricha-Alami$^{1,2}$}
%\email{ferrichaalami@yahoo.fr}
%\author{Z. Sakhi$^{1,2}$}
%\email{zb.sakhi@gmail.com}
%\author{M. Bennai$^{1,2}$}
%\email{mdbennai@yahoo.fr}

\affiliation{$^{1}${\small Quantum Physics and Magnetism Team, LPMC, Faculty of Science
Ben M'sik,}\\
{\small Casablanca Hassan II University, Morocco}\\} 
\affiliation{$^{2}$
Centre of Physics and Mathematics, CPM, Rabat, Morocco\\}

\maketitle

\section{Introduction}

%%%%%%%%%%%%%%%%%%%%%%%%%%%%%%%%%%%%%%%%%%%%%%

Affleck-Dine baryogenesis \cite{I1,I2} is a notable method for explaining
the baryon asymmetry in the universe. This mechanism involves the
utilization of a scalar field that possesses baryon number. Within the
framework of the Minimal Supersymmetric Standard Model (MSSM), there exist
numerous scalar fields known as flat directions, which are composed of
squarks (and sleptons). In the limit of supersymmetry (SUSY), the scalar
potential associated with these flat directions becomes zero, but it is
altered by the effects of SUSY breaking and higher order operators \cite{I3}%
. Within the Minimal Supersymmetric Standard Model, there exist numerous
flat directions that lack classical potentials. These flat directions are
only subject to modifications through soft supersymmetry breaking mass terms
and non-renormalizable terms. The presence of these flat directions gives
rise to several intriguing consequences in cosmology, including baryogenesis
and Q-ball formation. In the Affleck-Dine (AD) mechanism for baryogenesis, a
complex field associated with a flat direction (referred to as the AD field)
possesses a large value during the period of inflation. Subsequently, as the
mass of the AD field exceeds the Hubble parameter H, it begins to oscillate.
During this time, the AD field acquires a velocity in the phase direction
due to the presence of a baryon number-violating operator, leading to an
efficient generation of baryon number \cite{I4}.

In the context of primordial inflation \cite%
{I5,i7,i8,i9,i10,i11,i12,i13,i14,i15}, scalar fields characterized by
shallow potential energy can achieve non-zero vacuum expectation values
(VEVs) \cite{I6,o1,o2,o3,o4,o5,o6,o7}, supersymmetric extensions of the
standard model incorporate numerous scalar fields known as flat directions,
which exhibit a vanishing scalar potential under the condition of exact
supersymmetry. D-term potential is employed in the construction of
supersymmetric inflation models. Within these D-term inflation models,
supergravity effects do not introduce mass terms of the order of $O(H^{2})$,
where $H$ is the Hubble parameter. Instead, the AD field in these models
possesses only a soft supersymmetry-breaking mass term at the scale of the
weak force. Consequently, the AD field can acquire a substantial expectation
value during inflation, facilitating the success of AD baryogenesis \cite%
{I7,I8,I9,I10}.

The nature of dark matter (DM) remains a fundamental inquiry in cosmology,
particularly regarding the particle physics origin of the dark matter
particle \cite{I11}. Within the framework of the Minimal Supersymmetric
Standard Model and its extensions, the identity of the dark matter candidate
is intrinsically linked to the nature of supersymmetry breaking \cite{I12}.
In scenarios of gravity-mediated SUSY breaking, the neutralino emerges as a
natural candidate for the lightest SUSY particle (LSP). Conversely, in
gauge-mediated SUSY breaking (GMSB), the gravitino assumes the role of a
natural LSP candidate \cite{I13}. In GMSB (gauge-mediated supersymmetry
breaking) models, there are two potential mechanisms for the production of a
gravitino dark matter density. Firstly, if the reheating temperature $T_{re}$
is sufficiently high (around $10^{7}GeV$) and the gravitino mass ($m_{3/2}$)
is approximately $0.1GeV$, gravitinos can be generated through thermal
scattering \cite{I14}. Alternatively, gravitinos may be produced through the
late decay of next-to-lightest supersymmetric particles (NSPs). However,
this latter process is subject to strong constraints imposed by big-bang
nucleosynthesis (BBN) \cite{I15}.

The formation of Q balls during the Affleck-Dine baryogenesis process is a
widely recognized phenomenon \cite{I16,I17}. Notably, in the context of
gauge-mediated supersymmetry breaking, Q balls with sufficiently large
charges (Q) exhibit stability against decay into baryons (nucleons).
Consequently, these Q balls have the potential to serve as dark matter in
the universe and may be detectable. The baryon asymmetry observed in the
universe is attributed to the baryon numbers that are not retained within
the Q balls \cite{I18}. In the realm of scalar field theory, a Q-ball refers
to a non-topological soliton that possesses a conserved global charge \cite%
{I19}. A Q-ball solution arises when the energy minimum occurs at a non-zero
value of the scalar field while maintaining a fixed charge Q. During its
rotation, the condensate of the AD field experiences spatial instabilities,
leading to the formation of spherical lumps known as Q balls \cite{I20}. Q
balls charge essentially corresponds to the baryon number associated with
the AD field \cite{I21}. The study of the formation of large Q-balls has
primarily been conducted through analytical approaches employing linear
theory \cite{C8,C9}, as well as numerical simulations conducted on
one-dimensional lattices \cite{I16}. However, both methods rely on the
assumption that the Q-ball configuration is spherical, which prevents us
from accurately determining whether the Q-ball configuration is truly
achieved. In this work, we consider the AD mechanism in light of the D-term
inflation to discuss the observed baryon asymmetry that lies in AD
baryogenesis. The recent formalism that aims to derive the reheating
temperature as a function of the inflationary parameters to constrain the
reheating temperature according to Planck's data, relating the baryon
asymmetry to the reheating temperature provides an alternative way to study
the observed value of $n_{B}/s,$ NSPs abundance\ and the gravitino dark
matter mass according to observation. The paper is organized as follows: In
Sec. \ref{sec2} we present the Affleck-Dine mechanism and derive the general
form of the baryon number density. In Sec. \ref{sec3} we discuss the D-term
inflation in the AD mechanism. In Sec. \ref{sec4} we provide the basic
formalism for deriving the reheating temperature in D-term inflation then
study the observational constraints on Baryogenesis. In Sec. \ref{sec5} we
study the Q-ball decay and the abundance of NSP to constrain gravitinos dark
matter. In Sec. \ref{sec6} we conclude.

\section{ Affleck--Dine Mechanism in MSSM}

\label{sec2}

Let's take a brief look at the Affleck-Dine mechanism \cite{A1,A2} within
the MSSM in this section. The MSSM's superpotential is presented in \cite{A3}%
.

\begin{equation}
W=\gamma _{U}QUH_{U}+\gamma _{D}QDH_{D}+\gamma _{E}LEH_{D}+\mu H_{U}H_{D},
\label{eq1}
\end{equation}%
The superfields $H_{U},H_{D},U,D,L,E,$\ and $Q$\ respectively represent
Higgs doublets, up-type and down-type right-handed quarks, left-handed
lepton doublets, right-handed charged leptons, and left-handed quark
doublets. The supersymmetric theories of the field space encompass numerous
directions where the D-term contributions to the scalar potential vanish
entirely. The scalar potential along these directions is derived solely from
F-terms and SUSY-breaking contributions. Among these "flat" directions,
there is a captivating category consisting of the SUSY partners of the SM
fermions, such as squarks, sleptons, and Higgs fields.

The AD field potential rise from soft SUSY breaking terms and
nonrenormalizable terms \cite{A4,A5}. Therefore, we present the
superpotential that elevates a level direction $\phi $ as%
\begin{equation}
W=\frac{\lambda _{n}}{nM^{n-3}}\phi ^{n}.  \label{eq2}
\end{equation}%
Here, $\phi $ denotes the superfield that encompasses the flat direction $%
\phi $ and its corresponding fermionic partner, while $M$ refers to a large
scale that functions as a cut-off, and $\lambda _{n}\sim O(1)$ denotes the
Yukawa coupling. One can express the scalar potential of the AD field along
a flat direction as \cite{A6} 
\begin{equation}
V(\phi )=\left( m_{\phi }^{2}-c_{H}H^{2}\right) \left\vert \phi \right\vert
^{2}+\frac{1}{M^{2n-6}}\phi ^{2n-2}+\frac{m_{3/2}}{nM^{n-3}}\left( a_{m}\phi
^{n}+h.c\right) .  \label{eq3}
\end{equation}%
The potential terms that are proportional to the soft mass squared $m_{\phi
}^{2}$ and the gravitino mass $m_{3/2}$ are a result of SUSY breaking at the
true vacuum. The Hubble parameter $H$ in the expression reflects the impact
of $SUSY$ breaking, triggered by the inflaton's finite energy density \cite%
{A2}. Here, $c_{H}$ denotes a real constant which approximately is of the
order unity. The calculation of the baryon asymmetry is relatively
straightforward. The baryon number density correlates with the AD field, as
follows:%
\begin{equation}
n_{B}=\beta i\left( \dot{\phi}\phi -\phi ^{\ast }\dot{\phi}^{\ast }\right) .
\label{eq4}
\end{equation}%
The AD field has a corresponding baryon charge $\beta $, which is limited to
a maximum of $1/3$. The equation describing the AD field's motion in an
expanding universe can lead to deriving the time evolution of the baryon
number density%
\begin{equation}
\dot{n}_{B}+3Hn_{B}=2\beta \frac{m_{3/2}}{M^{n-3}}Im\left( a_{m}\phi
^{n}\right) ,  \label{eq5}
\end{equation}%
the baryon comoving density is expressed as:%
\begin{equation}
n_{B}(t)=\frac{1}{a(t)^{3}}\int^{t}a(t^{\prime })^{3}2\beta \frac{\left\vert
a_{m}\right\vert m_{3/2}}{M^{n-3}}\sin \theta dt,  \label{eq6}
\end{equation}%
the generated baryon number density at the oscillation time $t_{os}$ \cite%
{A6} 
\begin{equation}
n_{B}(t_{os})=\frac{2\left( n-2\right) }{3\left( n-3\right) }\beta \delta
_{eff}\left\vert a_{m}\right\vert m_{3/2}\left( H_{os}M^{n-3}\right)
^{2/\left( n-2\right) },  \label{eq7}
\end{equation}

here $\delta _{eff}\equiv \sin \theta \left( =O(1)\right) .$ Upon completion
of the reheating process following inflation, this results in the following
baryon asymmetry: \ 
\begin{equation}
\frac{n_{B}}{s}=\frac{1}{4}\frac{T_{re}}{M_{p}^{2}H_{os}^{2}}n_{B}(t_{os}),
\label{eq8}
\end{equation}%
the equation incorporates the reheating temperature $T_{re}$, $s$ as the
entropy density of the universe, and $M_{p}=2.4\times 10^{18}GeV$ as the
reduced Planck scale. It can be expressed using $H_{os}\simeq m_{\phi }$ as 
\cite{I6}:%
\begin{equation}
\frac{n_{B}}{s}\approx 10^{-10}\times \beta \left( \frac{1TeV}{m_{\phi }}%
\right) \left( \frac{T_{re}}{10^{9}GeV}\right) ,  \label{eq9}
\end{equation}%
for $n=4,$\ and for the case $n=6$, 
\begin{equation}
\frac{n_{B}}{s}\approx 10^{-10}\times \beta \left( \frac{1TeV}{m_{\phi }}%
\right) ^{1/2}\left( \frac{T_{re}}{100GeV}\right) .  \label{eq10}
\end{equation}%
These predictions on the baryon asymmetry by the Affleck-Dine mechanism look
promising. However, the Affleck-Dine condensate tends to be vulnerable to
spatial perturbations, leading to the formation of\ $Q$-balls, which are
nontopological solitons. The presence of $Q$-ball formation does not change
the projections regarding the baryon asymmetry, provided that any generated $%
Q$-balls are unstable and disintegrate prior to the onset of big-bang
nucleosynthesis.

\section{Dynamics of the inflaton field in AD mechanism}

\label{sec3}

Taking into account the possibility that D-terms provide the dominant
contribution to the vacuum energy \cite{B1,B2,B3}, we consider D-term
inflation \cite{B4,B5} which incorporates a neutral chiral superfield, $S$,
and superfields $\psi _{\pm }$ possessing charges of $\pm 1$ under a $U(1)$
symmetry \cite{B6}.%
\begin{equation}
V=\left\vert \kappa \right\vert ^{2}\left( \left\vert \psi _{+}\psi
_{-}\right\vert ^{2}+\left\vert S\psi _{+}\right\vert ^{2}+\left\vert S\psi
_{-}\right\vert ^{2}\right) +\frac{g^{2}}{2}\left( \left\vert \psi
_{+}\right\vert ^{2}-\left\vert \psi _{-}\right\vert ^{2}+\xi ^{2}\right)
^{2}.  \label{eq11}
\end{equation}%
In this scenario, we have added a positive Fayet-Iliopoulos (FI) term, $\xi
^{2}$, and the potential's global minimum, which preserves supersymmetry,
occurs at $S=0$, $\psi _{+}=0$, and $\psi _{-}=\xi $. Additionally, there is
a local minimum at $\psi _{+}=\psi _{-}=0$ for $\left\vert S\right\vert
>S_{c}\equiv g\xi /\kappa $. By making a $U(1)_{FI}$ phase rotation, we can
assume that $S$ is real. Introducing $\sigma \equiv \sqrt{2}S$, the
potential of $\sigma $, including a 1-loop correction, is provided for $%
\sigma >\sigma _{c}=\sqrt{2}g\xi /\kappa $ by \cite{BQ}

\begin{equation}
V\left( \sigma \right) =\frac{g^{2}\xi ^{2}}{2}\left( 1+\alpha \ln \left( 
\frac{\kappa \sigma ^{2}}{\Lambda ^{2}}\right) \right) ,  \label{eq12}
\end{equation}%
here $\alpha =g^{2}/16\pi ^{2}$ and $\Lambda $ is a renormalization mass
scale that does not affect physical quantities. When the AD scalar $\phi $
has an initial value of $\phi _{i}\sim O(M)$, the potential is mainly
influenced by the F-term of the AD field. As a result of supergravity
effects, a mass term of $O(H^{2})$ arises for the inflaton, causing it to
quickly roll down to its actual minimum. Hence, inflation is not feasible
under these circumstances. To create a successful inflationary model, the
value of $\phi _{i}$ must be less than $\phi _{c}$, which is defined as \cite%
{B7}:%
\begin{equation}
\phi _{c}=\sqrt{2}\left( \frac{g}{\sqrt{2}\lambda }\xi ^{2}M^{n-3}\right) ^{%
\frac{1}{n-1}}.  \label{eq13}
\end{equation}%
For $\phi _{i}\leq \phi _{c}$, the universe is dominated by the D-term of
the inflaton field. It can be demonstrated simply that the slow roll
criterion for $\sigma $ is met with $\epsilon \ll \eta $. The slow roll
parameters are defined as $\epsilon =\left( M_{p}^{2}/2\right) \left(
V^{\prime }(\sigma )/V(\sigma )\right) ^{2}$\ and $\eta =M_{p}^{2}V^{\prime
\prime }(\sigma )/V(\sigma ),$ knowing that the period of inflation will
last for a specific e-folds number given as $N\simeq \frac{1}{M_{p}^{2}}%
\int_{\sigma _{end}}^{\sigma _{k}}\frac{V}{V^{\prime }}d\sigma $.

\section{Reheating and Baryogenesis}

\label{sec4}

\subsection{Reheating Mechanism}

Following the end of inflation, there needs to be a reheating stage where
the energy stored in the inflaton field is transformed into a plasma
comprising of relativistic particles. After this stage, the conventional
radiation-dominated evolution of the early universe takes over. The energy
from the vacuum that drove inflation gets converted into potential and
kinetic energy, which is then utilized by the condensates of fields
oscillating around their various minima. In the previously presented model,
the vacuum energy, represented by $V=g^{2}\xi ^{4}/2$, is currently being
transported by some combination of the $\psi _{-}$ and $S$ fields, in each
oscillating condensate the relative portion of the energy depends on their
effective masses after inflation, that are respectively given by $\lambda
\xi $\ and $\sqrt{2}g\xi \ $\cite{B6}. The universe maintains a low
temperature until $H$ falls below the decay width of either the $S$ or $\psi
_{-}$ fields. At this point, one of these fields will undergo decay,
releasing its accumulated energy and causing the universe to undergo a
process of reheating.\ 

During the initial stages, the quasi-de-Sitter phase for $N_{k}$ e-folds of
expansion is powered by the inflaton field. The scale of the comoving
horizon decreases proportionally to $\sim a^{-1}$. Once the accelerated
expansion ends and the comoving horizon begins to expand, the phase of
reheating commences. Following an additional $N_{re}$ e-folds of expansion,
all the energy contained within the inflaton field has been entirely
dissipated, leading to the formation of a hot plasma having a reheating
temperature of $T_{re}$. Subsequent to that phase, the Universe undergoes
radiation domination for an additional $N_{RD}$ e-folds of expansion before
its transition into a state of matter domination.

In cosmology, we detect perturbation modes that are of a similar magnitude
to that of the horizon. To illustrate, Planck determines the pivot scale at $%
k=0.05Mpc^{-1}$. The comoving Hubble scale $a_{k}H_{k}=k$ is linked to the
present time scale relative to when this mode exited the horizon in the
following form \cite{B8},%
\begin{equation}
\ln \left( \frac{k}{a_{0}H_{0}}\right) =\ln \left( \frac{a_{k}}{a_{end}}%
\right) +\ln \left( \frac{a_{end}}{a_{re}}\right) +\ln \left( \frac{a_{re}}{%
a_{eq}}\right) +\ln \left( \frac{a_{eq}H_{eq}}{a_{0}H_{0}}\right) +\ln
\left( \frac{H_{k}}{H_{eq}}\right) .  \label{eq14}
\end{equation}%
Quantities denoted by a subscript $\left( k\right) $ are computed at the
moment when the horizon is exited. Other subscripts correspond to different
eras such as the end of inflation $(end)$, reheating $(re)$,
radiation-matter equality $(eq)$, and the current time $(0)$. knowing that $%
\ln \left( \frac{a_{k}}{a_{end}}\right) =N_{k},$\ $\ln \left( \frac{a_{end}}{%
a_{re}}\right) =N_{re}$\ and $\ln \left( \frac{a_{re}}{a_{eq}}\right)
=N_{RD}.$\ 

Knowing that $g^{2}\xi ^{2}$\ dominates in Eq. (\ref{eq12}) which leads to $%
V^{\prime }/V=\alpha /\sigma $\ and $V^{\prime \prime }/V=-\alpha /\sigma
^{2},$\ the Hubble parameter during inflation is derived as%
\begin{equation}
H_{k}=\pi M_{p}\sqrt{\frac{12\alpha A_{s}(1-n_{s})}{(6+\alpha )}},
\label{eq15}
\end{equation}%
\ with the scalar amplitude $\ln (10^{10}A_{s})=3.089_{-0.027}^{+0.024}$\
according to recent results \cite{B9}.\ For specific values of $\alpha ,$\ $%
H_{k}$\ is determined from $(1-n_{s}).$ The inflaton field at the end of
inflation can be estimated considering $\left\vert \eta \right\vert
=\left\vert \eta _{end}\right\vert \simeq 1$\ will\ lead to 
\begin{equation}
\sigma _{end}=\frac{\sqrt{\alpha }M_{p}}{\eta _{end}},  \label{eq16}
\end{equation}%
while its value at the beginning of inflation satisfies 
\begin{equation}
\sigma _{k}=M_{p}\sqrt{\frac{\alpha (6+\alpha )}{3(1-n_{s})}}.  \label{eq17}
\end{equation}%
\ In this direction, one would drive the inflationary e-folds as 
\begin{equation}
N_{k}=\frac{\sigma _{k}^{2}}{2M_{p}^{2}},  \label{eq18}
\end{equation}%
and the potential at the end of inflation in the following way $%
V_{end}=V(\sigma _{end})$. In addition to Eq. (\ref{eq14}), a second
relation of the reheating e-folds can be derived by tracking the energy
density at the end of inflation and the reheating phase as,%
\begin{equation}
N_{re}=\frac{1}{3\left( 1+\omega _{re}\right) }\ln \left( \frac{\rho _{end}}{%
\rho _{re}}\right) ,  \label{eq19}
\end{equation}%
here $\omega _{re}$ is the equation of state\ corresponding to the reheating
phase, The reheating temperature is determined by the final energy density $%
\rho _{re}=\left( \pi ^{2}/30\right) g_{re}T_{re}^{4},$ with $g_{re}$\ is
the number of relativistic species at the end of reheating. On the other
hand, in Ref. \cite{B10} they estimated the energy density at the end of
inflation by $\rho _{end}=\left( 3/2\right) V_{end}.$\ Assuming that entropy
remains conserved from the end of reheating until the present time, one can
establish a relationship between the temperature at the end of reheating $%
T_{re}$ and the CMB temperature $T_{0}$ taking into account $N_{k}$ the
number of e-foldings during inflation, the Hubble parameter $H_{k}$\ and the
value of the potential at the end of inflation $V_{end}$\ \cite{B10,B11},

\begin{equation}
T_{re}=\left[ \left( \frac{43}{11g_{re}}\right) ^{\frac{1}{3}}\frac{%
a_{0}T_{0}}{k}H_{k}e^{-N_{k}}\left[ \frac{3^{2}\cdot 5V_{end}}{\pi ^{2}g_{re}%
}\right] ^{-\frac{1}{3\left( 1+\omega _{re}\right) }}\right] ^{\frac{3\left(
1+\omega _{re}\right) }{3\omega _{re}-1}}.  \label{eq20}
\end{equation}%
\ \ 

It can be observed that when $N_{re}$ is greater, $T_{re}$ becomes smaller
and conversely. This means that reheating occurs more rapidly and
effectively, resulting in higher temperatures, which aligns with
expectations.

\subsection{Observational Constraints on Baryogenesis}

The cosmic accelerating expansion is caused by the inflaon's energy
dominance during the first stages of the universe's evolution, which leads
to the dilution of other matter contents and their asymmetry that existed
before inflation \cite{B12}. Conversely, the abundance of the baryon
asymmetry after the Big Bang Nucleosynthesis has been confirmed as $%
n_{B}/s=9.2\times 10^{-11}$ through the observation of CMB \cite{B9} and
measurements of the primordial abundances of light elements \cite{B13}.

Following the end of inflation, the system progresses and experiences
changes in the overall energy density and Hubble constant until a time when
the supergravity components in the potential equation (\ref{eq3}) governing
the AD direction reach a level of equivalence with the $-H^{2}\left\vert
\phi \right\vert ^{2}$ term. When $H$ is approximately equal to $m_{3/2}$,
the AD field begins to oscillate near the minimum. It is crucial to note
that during this phase, all terms in the potential for $\phi $ have similar
magnitudes, including those that conserve and violate baryon (or lepton)
numbers. Furthermore, the evolution of the $\phi $ direction as it
oscillates around zero results in significant CP-violating phases. In these
circumstances, the AD mechanism results a baryon-to-entropy ratio which is
obtained in the previous section. 
\begin{figure}[tbp]
\centering
\includegraphics[width=16cm]{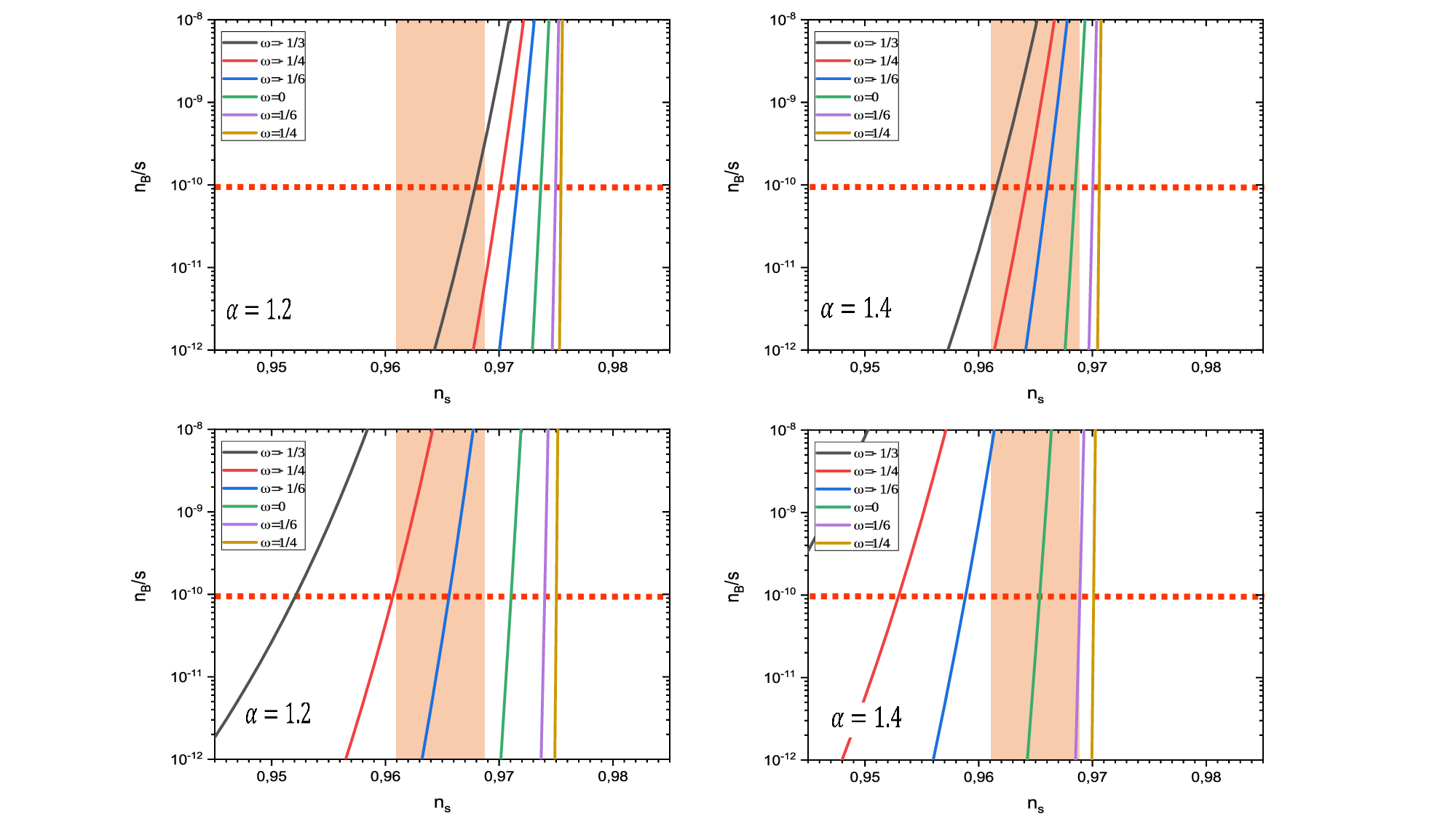}
\caption{{}Spectral index dependence of baryon number Considering baryon
asymmetry with both $n=4$ (top curves) and $n=6$ (bottom curves), we fixed
D-term potential parameter $\protect\alpha $\ to take $1.2$\ and $1.4$\
values. }
\label{fig:1}
\end{figure}

In Fig. (\ref{fig:1}) we present the behavior of baryon number as a function
of the spectral index. We study the $n_{B}/s$ evolution\ for some fixed
values of the equation of state parameter $\omega _{re}$\ and baryon
asymmetry number parameter $n$ corresponding to the AD mechanism, in the
case of the D-term inflationary potential considering both $\alpha =1.2,1.4.$
Note that the black line corresponds to $\omega _{re}=-1/3$, the red line
corresponds to $\omega _{re}=-1/4$, blue line corresponds to $\omega
_{re}=-1/6$, the green line corresponds to $\omega _{re}=0$, the purple line
corresponds to $\omega _{re}=1/6$ and the yellow line corresponds to $\omega
_{re}=1/4$. The vertical light orange region represents Planck-2018 bounds
on $n_{s}$\ and the red dashed line represents a precision of baryon number $%
\approx 10^{-10}$ from CMB experiments.\ We observe that for the case of $%
n=4 $ only lines with $\omega _{re}=-1/3$\ and $-1/4$\ tend towards the
central values of $n_{s}$\ for $n_{B}/s\lesssim 10^{-10}$ considering $%
\alpha =1.2,$\ while for $\alpha =1.4$\ the cases with $\omega _{re}=-1/6$\
and $-1/4$\ provide the best consistency with data. Howerver, for $\omega
_{re}=-1/3$\ and $0$ are partialy consistent for $n_{B}/s\gtrsim 10^{-10}$\
and $n_{B}/s\lesssim 10^{-10}$ respctevly. On another hand, when we consider 
$n=6$\ and $\alpha =1.2,$ the line with $\omega _{re}=-1/6$\ still provide
good consistency for all our chosen interval of $n_{B}/s$, while $\omega
_{re}=-1/4$\ presents fine compatibility only when $n_{B}/s\gtrsim 10^{-10}.$%
\ Moreover, $\omega _{re}=0$\ is the best compatible case for $\alpha =1.4.$
From this analysis we can conclude that the observed reheating temperature
from CMB data provides a wide range of possible values of the baryon
asymmetry number far away from the central value $n_{B}/s=9.2\times
10^{-11}, $ which affects the predicted mass of gravitino dark matter from
the AD mechanism as we will discuss in the next section.\ \ \ 

\section{Q-Balls and Dark Matter}

\label{sec5}

When the reheating temperature that occurred after inflation is as high as $%
10^{10}GeV$, the thermal processes can generate enough gravitinos to
potentially serve as a dark matter for a wide mass range in certain
supergravity models \cite{C1}. It is worth noting that the decay of the
next-to-lightest supersymmetric particle could also play a role in
generating the lightest supersymmetric particle gravitino abundance \cite%
{C2,C3}. A $Q$-ball is a type of coherent state of a complex scalar field
that exists and remains stable thanks to the conservation of a particular
global U(1) quantum number \cite{C4}. Within the MSSM framework, the
standard baryon and lepton numbers may serve as the conserved quantity for $%
Q $-balls composed of squarks and sleptons, respectively. It is worth
exploring whether these entities could have emerged in the early universe
and if their lifespan is sufficiently long to allow them to persist until
today and potentially contribute to dark matter \cite{I19,C6}. If $Q$-ball
formation takes place during the inflationary period before the reheating
process is finished, then the fluctuation will become non-linear once the
Hubble parameter reaches a certain threshold \cite{A6},%
\begin{equation}
H=H_{non}=\frac{m_{\phi }\left\vert K\right\vert }{2\alpha },  \label{eq21}
\end{equation}%
where $\alpha $\ is adopted in\ previous analyses to have values around $%
\alpha \simeq 30-40,$\ $K$ is the parameter involved in the one-loop
correction of the scalar potential of $\phi $ at the time of $Q$-ball
formation \cite{C8,C9}. At the time when the Hubble parameter reaches $%
H_{non}$, the baryon number density of the condensate reaches,%
\begin{equation}
n_{B}(t_{non})\simeq n_{B}(t_{osc})\times \left( \frac{H_{non}}{H_{osc}}%
\right) ^{2}.  \label{eq22}
\end{equation}%
It is possible to estimate the charge carried by an individual Q-ball as 
\cite{A6},%
\begin{eqnarray}
Q &\simeq &\frac{4}{3}\pi R_{Q}^{3}\times n_{B}(t_{non})  \notag \\
&\sim &3\times 10^{-3}\times \frac{2\left( n-2\right) }{3\left( n-3\right) }%
\beta \delta _{eff}\left\vert a_{m}\right\vert \left( \frac{m_{3/2}}{m_{\phi
}}\right) \left( \frac{M}{m_{\phi }}\right) ^{\frac{2\left( n-3\right) }{%
\left( n-2\right) }}  \label{eq23}
\end{eqnarray}

When a $Q$-ball interacts with particles in a thermal bath, a portion of its
charge may evaporate. The amount of charge that evaporates is predicted to
be around $\Delta Q=O(10^{18})$ \cite{C9}. Therefore, if the charge of a $Q$%
-ball is greater than $\Delta Q$, the $Q$-ball could potentially survive the
evaporation process. The abundance of NSP following the decay of Q-balls (up
until the decay of NSP) can be expressed as \cite{C10}:%
\begin{equation}
Y_{N}(T)\simeq \left[ \frac{1}{Y(T_{d})}+\sqrt{\frac{8\pi ^{2}g_{\ast
}(T_{d})}{45}}\left\langle \sigma \upsilon \right\rangle M_{p}\left(
T_{d}-T\right) \right] ^{-1}.  \label{eq24}
\end{equation}%
While if annihilation is not efficient after the production of NSPs via
Q-ball decay, the resulting abundance is expressed as \cite{C10}:%
\begin{equation}
Y_{N}=3\times 10^{-10}\left( \frac{N}{3}\right) \left( \frac{f_{B}}{1}%
\right) \left( \frac{n_{B}/s}{10^{-10}}\right) ,  \label{eq25}
\end{equation}%
we should bring to mind as well $\rho _{c}/s\simeq 1.7\times 10^{-9}\left(
h/0.7\right) ^{2}GeV.$ The notion that Affleck-Dine baryogenesis could
explain both the baryon asymmetry and dark matter through $Q$-ball decay was
abandoned in favor of neutralino in the MSSM \cite{C11}. In the case of $n=6$
Affleck-Dine baryogenesis, experimental evidence doesn't exclude the
possibility that dark matter particle mass must be of $1GeV$, it appears to
be somewhat unnatural and improbable in the context of gravity-mediated SUSY
breaking.

\begin{figure}[tbp]
\centering
\includegraphics[width=16cm]{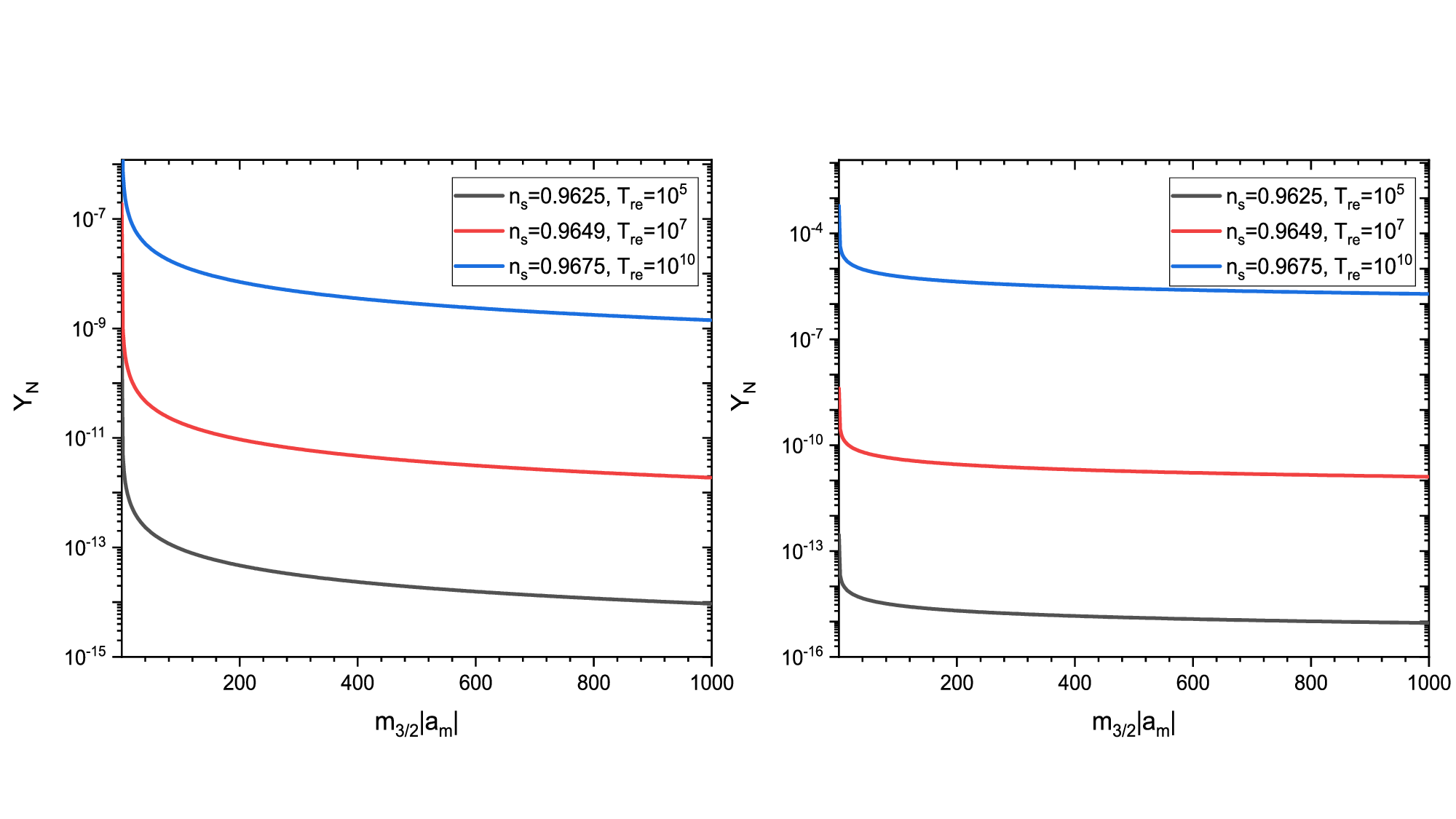}
\caption{{}Variation of the NSPs abundance as a function of the gravitino
mass considering different index spectral values and reheating temperature
in $GeV.$ The left plot corresponds to\ $n=4$, and the right plot
corresponds to $n=6$. }
\label{fig:2}
\end{figure}
Fig. (\ref{fig:2}) shows the evolution of the abundance of NSPs as a
function of gravitino mass knowing that $m_{\phi }\simeq m_{3/2}\left\vert
a_{m}\right\vert $ and $\left\vert a_{m}\right\vert \approx 10$,\ from the
figure\ we observe that the gravitinos mass has a weak effect on the
behavior of $Y_{N}.$\ However, increasing the reheating temperature causes a
decrease in the abundance of NSPs. On the other hand, for the case of $n=6,$
the NSPs abundance decreases quickly for $m_{3/2}\leq 20GeV,$ while for $n=6$
the abundance of NSPs stabilizes when $m_{3/2}\gtrsim 2GeV.$\ Moreover, the
most compatible value of reheating temperature is given\ by$\ T_{re}\simeq
10^{7}GeV$ since it coincides with the central value of $n_{s}=0.9649$ which
equivalently predicts values of NSPs abundance\ around $Y_{N}\simeq 10^{-11}$%
\ for both $n=4$ and $n=6$ that corresponds to the AD mechanism.\ When the
RH sneutrino NLSPs possess suitable characteristics, the decay of the $d=6$
Q-balls into RH sneutrinos could potentially generate gravitino dark matter
while remaining consistent with Big Bang Nucleosynthesis, even in cases
where the gravitino mass $m_{3/2}$ exceeds $1GeV$, while if the
fragmentation of the condensate results in only positively charged Q-balls,
the preservation of B-conservation and R-parity conservation indicates that
the gravitino mass should be approximately $2GeV$ \cite{Cx12}. To achieve a
successful scenario with RH sneutrino NLSPs in this mass range, two
conditions must be met: firstly, the MSSM-LSPs resulting from Q-ball decay
should be able to decay into RH sneutrinos before nucleosynthesis, and
secondly, the RH sneutrinos should be able to decay into gravitinos without
violating constraints imposed by nucleosynthesis and free-streaming \cite%
{Cx13}. This could be accomplished if the RH neutrinos possess enhanced
Yukawa couplings through the see-saw mechanism \cite{Cx14}.

When the annihilation takes place following the production of NSPs, the
resulting abundance of NSPs for temperatures significantly lower than $T_{d}$
can be expressed as follows:%
\begin{equation}
Y_{N}(T)\simeq \left[ \sqrt{\frac{8\pi ^{2}g_{\ast }(T_{d})}{45}}%
\left\langle \sigma \upsilon \right\rangle M_{p}T_{d}\right] ^{-1}.
\label{eq27}
\end{equation}%
Assuming this situation, the production of all dark matter gravitinos must
occur through the decay of NSPs. When the constraint on $Y_{N}$ is applied
based on the the equivalence of $\rho _{NSP},\rho _{LSP}$ and $\rho _{DM}$,
we can derive the following:

\begin{equation}
Y_{N}(T)\simeq 3\times 10^{-12}\left( \frac{3\times 10^{-8}GeV^{2}}{%
\left\langle \sigma \upsilon \right\rangle }\right) \left( \frac{1GeV}{T_{d}}%
\right) \left( \frac{10}{g_{\ast }(T_{d})}\right) ^{1/2}\sim 4\times
10^{-12}\left( \frac{100GeV}{m_{DM}}\right) ,  \label{eq28}
\end{equation}%
to achieve $T_{d}\sim 1GeV$, a substantial annihilation cross-section of h $%
\left\langle \sigma \upsilon \right\rangle \simeq 10^{-8}-10^{-7}GeV^{2}$ is
necessary, where $m_{DM}$ represents the mass of the dark matter particle.
For such a high level of annihilation cross-section, the neutralino NSP must
not be Bino-like but rather Higgsino-like \cite{C1}. When $T_{re}$ is less
than approximately $10^{7}GeV$, $Q$-balls possessing baryon numbers ranging
from $10^{12}$ to $10^{18}$ will decay subsequent to reheating and maintain
the effective Enqvist-McDonald correlation that links the energy densities
of baryonic matter and dark matter \cite{C12,C13},%
\begin{equation}
\left( \frac{\Omega _{N}h^{2}}{0.11}\right) \approx \left( \frac{m_{NSP}}{%
1.5GeV}\right) \left( \frac{Nf_{B}}{3}\right) \left( \frac{\Omega _{b}h^{2}}{%
0.02}\right) .  \label{eq29}
\end{equation}

\begin{figure}[tbp]
\centering
\includegraphics[width=12cm]{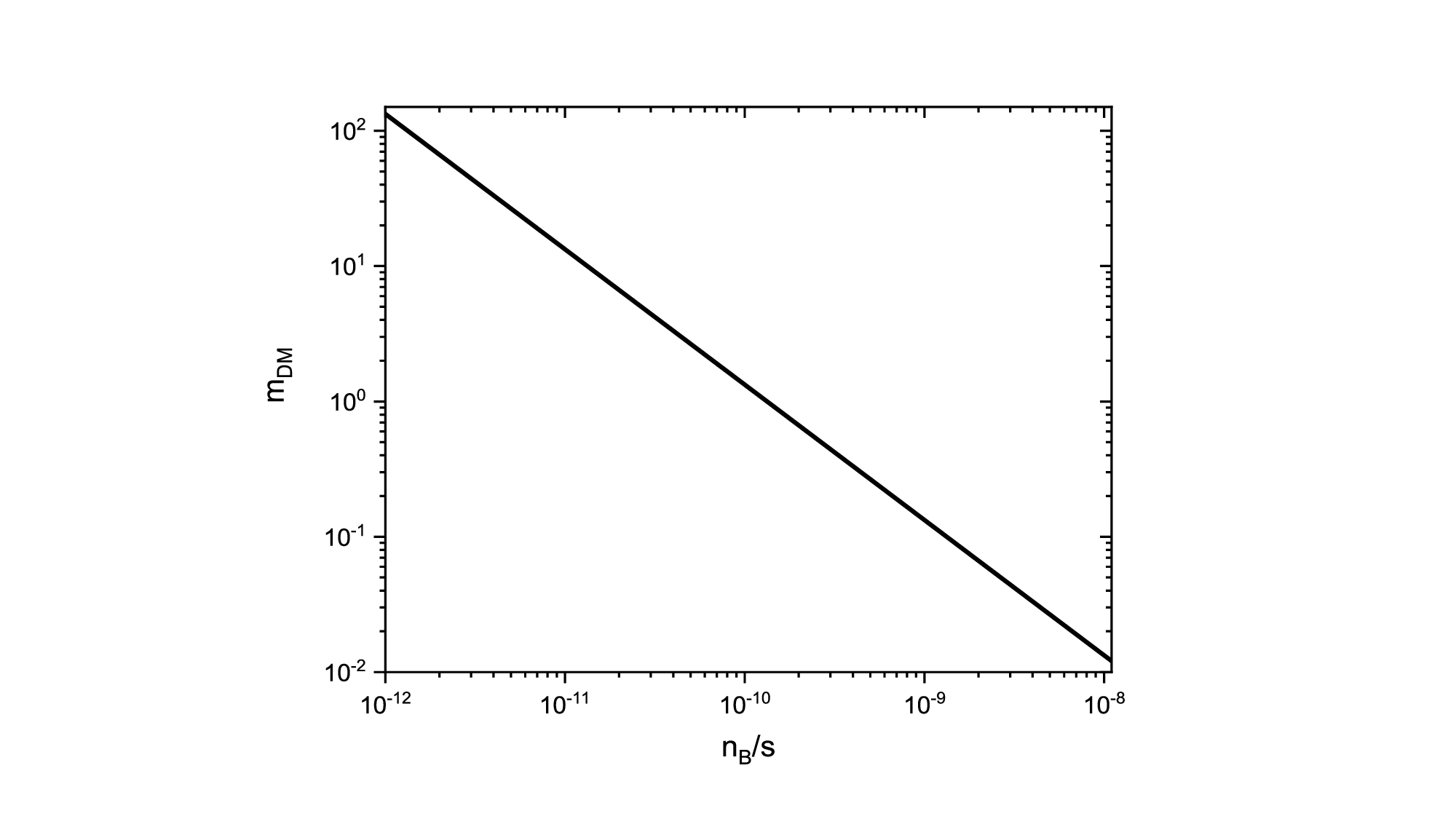}
\caption{{}Dark matter mass variation as a function of the baryon asymmetry
number.}
\label{fig:3}
\end{figure}
We study the behavior of the dark matter mass as a function of the baryon
asymmetry in Fig. (\ref{fig:3}), the mass of dark matter is decreased when
we consider higher values of $n_{B}/s.$\ From the predicted value of baryon
asymmetry, one would estimate dark matter masses around $m_{DM}\simeq 1GeV.$%
\ However, our analysis proves that the reheating temperature has a
noticeable consequence on the baryon asymmetry which can slightly modify the
range of dark matter rang for consistency with a wider number of models. \ 

\begin{figure}[tbp]
\centering
\includegraphics[width=12cm]{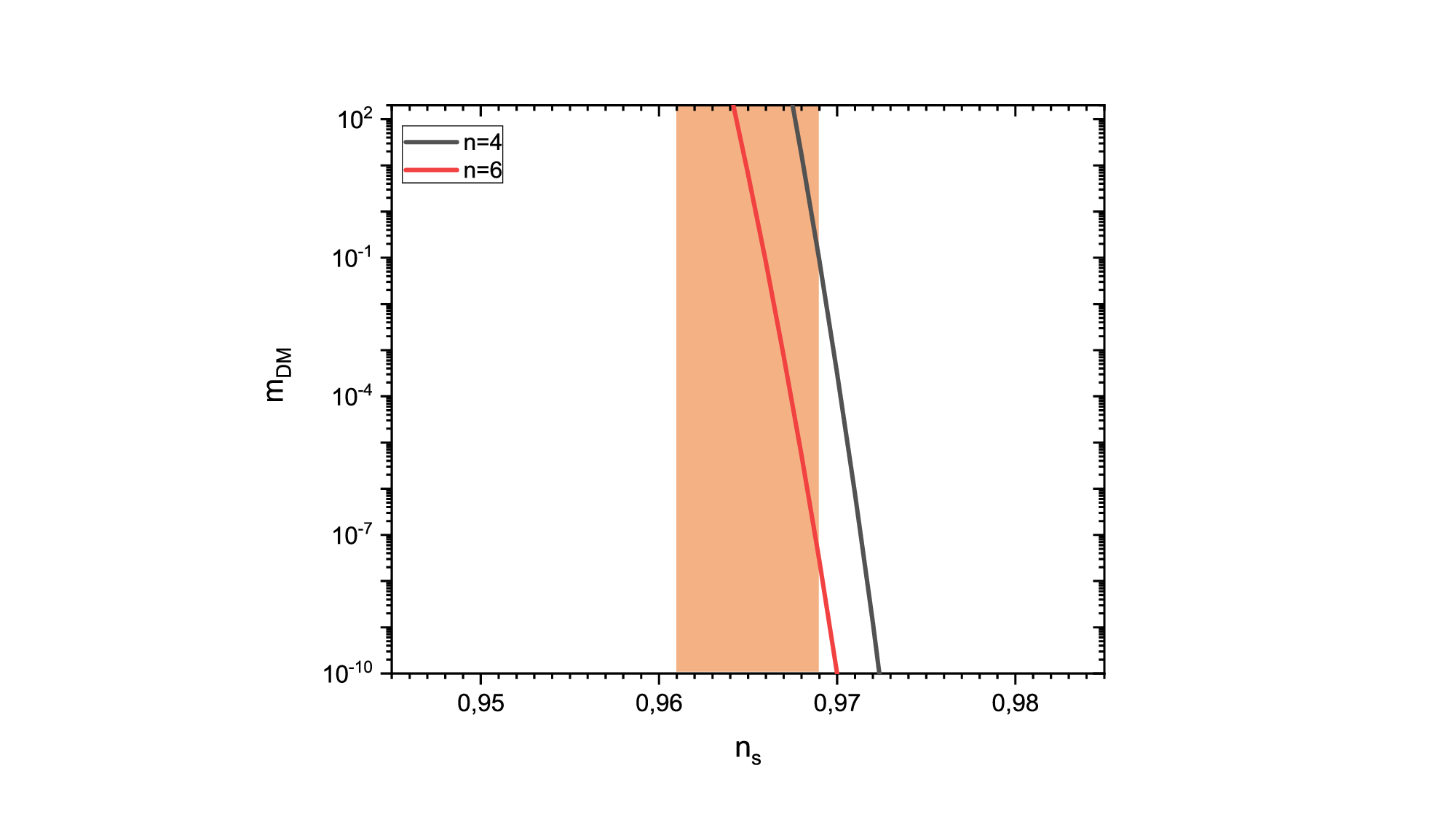}
\caption{{}Variation of the dark matter mass as a function of spectral index 
$n_{s}$, the black line corresponds to $n=4$\ and the red line corresponds
to $n=6.$ }
\label{fig:4}
\end{figure}
Fig. (\ref{fig:4}) represents the evolution of the dark matter mass with
respect to the spectral index, the central orange region corresponds to the
observed value of spectral index $n_{s}=0.9649\pm 0.0042.$ From the $%
m_{DM}-n_{s}$ plane we observe that CMB experiments predict lower bounds on
the dark matter masses, for the $n=4$ case dark matter masses are predicted
to be $m_{DM}\gtrsim 100MeV$, while for $n=6$ the lower bound on the\ dark
matter masses is given as $m_{DM}\gtrsim 100eV$. For both AD\ model cases,
even if the dark matter can reach a higher range of masses\ according to
recent Plancks data, the gravitino dark matter mass arising from NSP decay
must be below $1GeV$ to satisfy the constraints imposed by BBN on NSP decay.
The least restrictive upper bounds on BBN are obtained for stau and
sneutrino NSPs, which permit gravitino masses ranging from $0.1$ to $1GeV$.

\section{Conclusion}

\label{sec6}

The estimation of baryon asymmetry in recent advancements relies on whether
the AD field experiences early oscillation due to thermal effects at a high
reheating temperature. Furthermore, achieving the appropriate baryon
asymmetry is possible through the reduction of baryon asymmetry with the aid
of early oscillation. Additionally, Affleck-Dine baryogenesis remains a
feasible and compatible explanation for both gravitino dark matter and
baryon asymmetry, with the majority of gravitinos being generated through
NSP decay. Consequently, the nature of the NSP and the constraints imposed
by BBN and CMB play a crucial role in determining the conclusions related to
NSP decay, which is an important subject of study.

In summary, the previously proposed scenario involving Affleck-Dine
baryogenesis was reexamined with a focus on the abundance of Q-balls in
connection to gravitino dark matter. If Q-balls cannot withstand
evaporation, the production of gravitinos becomes reliant on the decay of
NSPs, with their abundance determined by the thermal relic density. However,
if Q-balls can survive evaporation, NSPs are generated as decay products of
Q-balls. Consequently, these persistent Q-balls offer a new potential avenue
for gravitino dark matter through NSP decay.

In this work, we have shown that the baryon asymmetry can be alternatively
constrained from plancks data using recent formalism that was proposed to
study the reheating phase. Moreover, the derived baryon number can provide a
good tool to discuss the NSP abundance from Q-balls decay. On another hand,
one can estimate gravitino dark matter by taking into account previous
findings on NSP decay.

In conclusion, our findings indicate that both baryon asymmetry and
gravitino dark matter can arise from the decay of Q-balls within the
gauge-mediated MSSM framework, while satisfying the constraints imposed by
BBN. In order to advance the model, a comprehensive numerical investigation
of Affleck-Dine condensate fragmentation and Q-ball decay is required. This
area of study will be the focus of forthcoming research efforts.


\begin{thebibliography}{99}
\bibitem{I1} Affleck, I., \& Dine, M. (1985). \textit{A new mechanism for
baryogenesis.} Nuclear Physics B, 249(2), 361-380.

\bibitem{I2} Dine, M., Randall, L., \& Thomas, S. (1996). \textit{%
Baryogenesis from flat directions of the supersymmetric standard model.}
Nuclear Physics B, 458(1-2), 291-323.

\bibitem{I3} Kasuya, S., \& Kawasaki, M. (2014). \textit{Q-ball dark matter
and baryogenesis in high-scale inflation.} Physics Letters B, 739, 174-179.

\bibitem{I4} Kawasaki, M., \& Takahashi, F. (2001). \textit{Adiabatic and
isocurvature fluctuations of Affleck--Dine field in D-term inflation model.}
Physics Letters B, 516(3-4), 388-394.

\bibitem{I5} Liddle, A. R. (1998). \textit{An introduction to cosmological
inflation}. High energy physics and cosmology, 260.

\bibitem{i7} El Bourakadi, K., Ferricha-Alami, M., Filali, H., Sakhi, Z., \&
Bennai, M. (2021). \textit{Gravitational waves from preheating in
Gauss--Bonnet inflation.} The European Physical Journal C, 81(12), 1144.

\bibitem{i8} El Bourakadi, K., Bousder, M., Sakhi, Z., \& Bennai, M. (2021). 
\textit{Preheating and reheating constraints in supersymmetric braneworld
inflation}. The European Physical Journal Plus, 136(8), 1-19.

\bibitem{i9} El Bourakadi, K., Asfour, B., Sakhi, Z., Bennai, M., \& Ouali,
T. (2022). \textit{Primordial black holes and gravitational waves in
teleparallel Gravity}. The European Physical Journal C, 82(9), 792.

\bibitem{i10} Sakhi, Z., El Bourakadi, K., Safsafi, A., Ferricha-Alami, M.,
Chakir, H., \& Bennai, M. (2020). \textit{Effect of brane tension on
reheating parameters in small field inflation according to Planck-2018 data.}
International Journal of Modern Physics A, 35(30), 2050191.

\bibitem{i11} K. El Bourakadi, Z. Sakhi and M. Bennai, \textit{Preheating
constraints in $\alpha$-attractor inflation and gravitational waves
production}, Int. J. Mod. Phys. A 37, 2250117 (2022).

\bibitem{i12} Bourakadi, K. E., Koussour, M., Otalora, G., Bennai, M., \&
Ouali, T. (2023). \textit{Constant-roll and primordial black holes in f (Q,
T) gravity.} arXiv preprint arXiv:2301.03696. Physics of the Dark Universe
(2023).

\bibitem{i13} Bourakadi, K. E., Sakhi, Z., \& Bennai, M. (2023). \textit{%
Observational constraints on Tachyon inflation and reheating in f (Q)
gravity.} arXiv preprint arXiv:2302.11229.

\bibitem{i14} Bourakadi, K. E., Sakhi, Z., \& Bennai, M. (2023). \textit{%
Cosmological Signatures from the Post-Inflationary Epochs.} arXiv preprint
arXiv:2303.04875.

\bibitem{i15} Bourakadi, K. E. (2023). \textit{Exploring the impact of Tree
Level Higgs Potential on Reheating in f (Q) Gravity}. arXiv preprint
arXiv:2303.02517.

\bibitem{I6} Allahverdi, R., \& Mazumdar, A. (2012). \textit{A mini review
on Affleck--Dine baryogenesis}. New Journal of Physics, 14(12), 125013.

\bibitem{o1} Ferricha-Alami, et al. \textit{Mutated hybrid inflation on
brane and reheating temperature}. The European Physical Journal Plus, 2017,
vol. 132, p. 1-10.

\bibitem{o2} Ferricha-Alami, M., et al. \textit{Smooth hybrid inflation on
brane constrained by Planck data}. International Journal of Modern Physics A
29.26 (2014): 1450146.

\bibitem{o3} Ferricha-Alami, M., Safsafi, A., Bouaouda, A., Zarrouki, R., \&
Bennai, M. (2015). \textit{K\"{a}hler potential braneworld inflation in
supergravity after Planck 2015}. International Journal of Modern Physics A,
30(34), 1550208.

\bibitem{o4} Ferricha-Alami, M., Mounzi, Z., Jdair, O., Naciri, M., Bennai,
M., \& Chakir, H. (2017). \textit{Randall--Sundrum II model from small field
inflation in light of planck data and reheating temperature}. Moscow
University Physics Bulletin, 72, 425-432.

\bibitem{o5} Ferricha-Alami, M., Jdair, O., Chakir, H., \& Bennai, M.
(2018). \textit{Logarithmic Potential Braneworld in Light of the Recent
Experiment Observation}. Revista Cubana de F\'{\i}sica, 35(2), 86-90.

\bibitem{o6} Mounzi, Z., Ferricha-Alami, M., Safsafi, A., \& Bennai, M.
(2016). \textit{Braneworld Inflation in Supergravity with a Shift Symmetric K%
\"{a}hler Potential}. Journal of Astrophysics and Astronomy, 37, 1-10.

\bibitem{o7} Mounzi, Z., Ferricha-Alami, M., Safsafi, A., \& Bennai, M.
(2017). \textit{Observational constraints on a hyperbolic potential in
brane-world inflation}. Gravitation and Cosmology, 23, 84-89.

\bibitem{I7} Gaillard, M. K., Murayama, H., \& Olive, K. A. (1995). \textit{%
Preserving flat directions during inflation.} Physics Letters B, 355(1-2),
71-77.

\bibitem{I8} Odintsov, S. D., \& Oikonomou, V. K. (2016). \textit{%
Gauss--Bonnet gravitational baryogenesis.} Physics Letters B, 760, 259-262.

\bibitem{I9} Oikonomou, V. K., Pan, S., \& Nunes, R. C. (2017). \textit{%
Gravitational baryogenesis in running vacuum models}. International Journal
of Modern Physics A, 32(22), 1750129.

\bibitem{I10} Odintsov, S. D., \& Oikonomou, V. K. (2016). \textit{Loop
quantum cosmology gravitational baryogenesis.} Europhysics Letters, 116(4),
49001.

\bibitem{I11} Feng, J. L. (2010). \textit{Dark matter candidates from
particle physics and methods of detection.} Annual Review of Astronomy and
Astrophysics, 48, 495-545.

\bibitem{I12} Bertone, G., Hooper, D., \& Silk, J. (2005). \textit{Particle
dark matter: Evidence, candidates and constraints}. Physics reports,
405(5-6), 279-390.

\bibitem{I13} Fujii, M., \& Yanagida, T. (2002). \textit{Natural gravitino
dark matter and thermal leptogenesis in gauge-mediated
supersymmetry-breaking models.} Physics Letters B, 549(3-4), 273-283.

\bibitem{I14} G\'{o}mez-Vargas, G. A., L\'{o}pez-Fogliani, D. E., Mu\~{n}oz,
C., \& Perez, A. D. (2020). \textit{MeV-GeV }$\mathit{\gamma }$\textit{-ray
telescopes probing axino LSP/gravitino NLSP as dark matter in the }$\mathit{%
\mu \upsilon }$\textit{SSM}. Journal of Cosmology and Astroparticle Physics,
2020(01), 058.

\bibitem{I15} Kawasaki, M., Kohri, K., Moroi, T., \& Yotsuyanagi, A. (2008). 
\textit{Big-bang nucleosynthesis and gravitinos.} Physical Review D, 78(6),
065011.

\bibitem{I16} Kusenko, A., \& Shaposhnikov, M. (1998). \textit{%
Supersymmetric Q-balls as dark matter}. Physics Letters B, 418(1-2), 46-54.

\bibitem{I17} Kasuya, S., \& Kawasaki, M. (2000). \textit{Q-ball formation
through the Affleck-Dine mechanism}. Physical Review D, 61(4), 041301.

\bibitem{I18} Kasuya, S., \& Kawasaki, M. (2014). \textit{Q-ball dark matter
and baryogenesis in high-scale inflation}. Physics Letters B, 739, 174-179.

\bibitem{I19} Coleman, S. (1985). \textit{Q-balls}. Nuclear Physics B,
262(2), 263-283.

\bibitem{I20} Kusenko, A., \& Shaposhnikov, M. (1998). \textit{%
Supersymmetric Q-balls as dark matter}. Physics Letters B, 418(1-2), 46-54.

\bibitem{I21} Kasuya, S., \& Kawasaki, M. (2014). \textit{Baryogenesis from
the gauge-mediation type Q-ball and the new type of Q-ball as the dark matter%
}. Physical Review D, 89(10), 103534.

\bibitem{C8} Enqvist, K., \& McDonald, J. (1998). \textit{Q-balls and
baryogenesis in the MSSM.} Physics Letters B, 425(3-4), 309-321.

\bibitem{C9} Enqvist, K., \& McDonald, J. (1999). \textit{B-ball
baryogenesis and the baryon to dark matter ratio.} Nuclear Physics B,
538(1-2), 321-350.

\bibitem{A1} Affleck, I., \& Dine, M. (1985). \textit{A new mechanism for
baryogenesis.} Nuclear Physics B, 249(2), 361-380.

\bibitem{A2} Dine, M., Randall, L., \& Thomas, S. (1995). \textit{%
Supersymmetry breaking in the early universe.} Physical Review Letters,
75(3), 398.

\bibitem{A3} Nilles, H. P. (1984). \textit{Supersymmetry, supergravity and
particle physics.} Physics Reports, 110(1-2), 1-162.

\bibitem{A4} Ng, K. W. (1989). \textit{Residual F-terms and the Affleck-Dine
mecahnism for baryogenesis.} Nuclear Physics B, 321(2), 528-540.

\bibitem{A5} Dine, M., Randall, L., \& Thomas, S. (1996). \textit{%
Baryogenesis from flat directions of the supersymmetric standard model.}
Nuclear Physics B, 458(1-2), 291-323.

\bibitem{A6} Fujii, M., \& Hamaguchi, K. (2002). \textit{Nonthermal dark
matter via Affleck-Dine baryogenesis and its detection possibility.}
Physical Review D, 66(8), 083501.

\bibitem{B1} Casas, J. A., \& Munoz, C. (1989). \textit{Inflation from
superstrings.} Physics Letters B, 216(1-2), 37-40.

\bibitem{B2} Casas, J. A., Moreno, J. M., Munoz, C., \& Quir\'{o}s, M.
(1989). \textit{Cosmological implications of an anomalous }$\mathit{U\ (1)}$%
\textit{: inflation, cosmic strings and constraints on superstring
parameters.} Nuclear Physics B, 328(1), 272-291.

\bibitem{B3} Stewart, E. D. (1995). \textit{Inflation, supergravity, and
superstrings}. Physical Review D, 51(12), 6847.

\bibitem{B4} Halyo, E. (1996). \textit{Hybrid inflation from supergravity
D-terms}. Physics Letters B, 387(1), 43-47.

\bibitem{B5} Bin\'{e}truy, P., \& Dvali, G. (1996). \textit{D-term inflation}%
. Physics Letters B, 388(2), 241-246.

\bibitem{B6} Kolda, C., \& March-Russell, J. (1999). \textit{Supersymmetric
D-term inflation, reheating, and Affleck-Dine baryogenesis}. Physical Review
D, 60(2), 023504.

\bibitem{BQ} Panotopoulos, G. (2005). \textit{D-term inflation in D-brane
cosmology}. Physics Letters B, 623(3-4), 185-191.

\bibitem{B7} Kawasaki, M., \& Takahashi, F. (2001). \textit{Adiabatic and
isocurvature fluctuations of Affleck--Dine field in D-term inflation model}.
Physics Letters B, 516(3-4), 388-394.

\bibitem{B8} Liddle, A. R., \& Leach, S. M. (2003). \textit{How long before
the end of inflation were observable perturbations produced?}. Physical
Review D, 68(10), 103503.

\bibitem{B9} Aghanim, N., Akrami, Y., Ashdown, M., Aumont, J., Baccigalupi,
C., Ballardini, M., ... \& Roudier, G. (2020). \textit{Planck 2018 results-VI%
}. Cosmological parameters. Astronomy \& Astrophysics, 641, A6.

\bibitem{B10} Cook, J. L., Dimastrogiovanni, E., Easson, D. A., \& Krauss,
L. M. (2015). \textit{Reheating predictions in single field inflation.}
Journal of Cosmology and Astroparticle Physics, 2015(04), 047.

\bibitem{B11} Dai, L., Kamionkowski, M., \& Wang, J. (2014). \textit{%
Reheating constraints to inflationary models}. Physical review letters,
113(4), 041302.

\bibitem{B12} Takeda, N. (2015). \textit{Inflatonic baryogenesis with large
tensor mode.} Physics Letters B, 746, 368-371.

\bibitem{B13} Cooke, R. J., Pettini, M., Jorgenson, R. A., Murphy, M. T., \&
Steidel, C. C. (2014). \textit{Precision measures of the primordial
abundance of deuterium.} The Astrophysical Journal, 781(1), 31.

\bibitem{C1} Bolz, M., Buchm\"{u}ller, W., \& Pl\"{u}macher, M. (1998). 
\textit{Baryon asymmetry and dark matter.} Physics Letters B, 443(1-4),
209-213.

\bibitem{C2} Feng, J. L., Rajaraman, A., \& Takayama, F. (2003). \textit{%
Superweakly interacting massive particles.} Physical review letters, 91(1),
011302.

\bibitem{C3} Feng, J. L., Su, S., \& Takayama, F. (2004). \textit{%
Supergravity with a gravitino lightest supersymmetric particle.} Physical
Review D, 70(7), 075019.

\bibitem{C4} Kusenko, A., \& Shaposhnikov, M. (1998). \textit{Supersymmetric
Q-balls as dark matter.} Physics Letters B, 418(1-2), 46-54.

\bibitem{C6} Kusenko, A. (1997). \textit{Solitons in the supersymmetric
extensions of the standard model.} Physics Letters B, 405(1-2), 108-113.

\bibitem{C10} Seto, O. (2006). \textit{Affleck-Dine baryogenesis and
gravitino dark matter.} Physical Review D, 73(4), 043509.

\bibitem{C11} Enqvist, K., \& McDonald, J. (2000). \textit{The dynamics of
Affleck--Dine condensate collapse.} Nuclear physics B, 570(1-2), 407-422.

\bibitem{Cx12} Doddato, F., \& McDonald, J. (2012). \textit{New Q-ball
solutions in gauge-mediation, Affleck-Dine baryogenesis and gravitino dark
matter.} Journal of Cosmology and Astroparticle Physics, 2012(06), 031.

\bibitem{Cx13} Ishiwata, K., Matsumoto, S., \& Moroi, T. (2008). \textit{%
Gravitino dark matter with weak-scale right-handed sneutrino.} Physical
Review D, 77(3), 035004.

\bibitem{Cx14} Doddato, F., \& McDonald, J. (2011). Comment on \textit{%
Gravitino Dark Matter and Baryon Asymmetry from Q-ball Decay in Gauge
Mediation (arXiv: 1107.0403)}. arXiv preprint arXiv:1107.1402.

\bibitem{C12} Shoemaker, I. M., \& Kusenko, A. (2009). \textit{Gravitino
dark matter from Q-ball decays}. Physical Review D, 80(7), 075021.

\bibitem{C13} Roszkowski, L., \& Seto, O. (2007). \textit{Axino dark matter
from }$Q$\textit{-balls in Affleck-Dine baryogenesis and the }$\Omega _{b}$%
\textit{-}$\Omega _{DM}$\textit{\ coincidence problem.} Physical review
letters, 98(16), 161304.
\end{thebibliography}
\end{document}